\documentclass[11pt]{article}
\usepackage{graphicx}

\def\FIG #1 #2 [#3] #4\par{%
  \begin{figure}[!h] \begin{center}%
    \includegraphics*[#3]{#2}%
    \caption{\label{#1}#4} 
  \end{center}\end{figure}%
}

\def\FIGG #1 #2 #3 [#4] #5\par{%
  \begin{figure}[!h]
    \includegraphics*[#4]{#2}
    \hfill
    \includegraphics*[#4]{#3}
    \caption{\label{#1}#5}
  \end{figure}
}

\title{Approximations to path integrals\\
         and spectra of quantum systems\thanks{Preprint ITEP-111-82}}%

\author{S.I.Blinnikov$^1$ and N.V.Nikitin$^2$ \\
   $^1$ITEP, 117218, Moscow, Russia \\
e-mail {\tt sergei.blinnikov@itep.ru} \\
$^2$MEPhI, 115409, Moscow, Russia \\ 
e-mail {\tt uniar@aha.ru} }

\date{1982 July 19}%
\begin{document}

\begin{titlepage}
\maketitle
\thispagestyle{empty}

\begin{abstract} An expression, suitable for practical calculations, for the Green
function $G(E;x_1,x_2)$ of the Schr\"{o}dinger equation is obtained through the
approximations of the path integral by $n$-fold multiple integrals. The
approximations to $\Re\{G(E;x,x)\}$ on the real $E$-axis have peaks near the
values of the energy levels $E_{j}$, where exact $\Re\{G(E;x,x)\}$ behaves like
$\delta(E-E_j)$. The analytic and numerical examples for one-dimensional and
multi-dimensional harmonic and anharmonic oscillators, and potential wells, show
that median values of these peaks for approximate $G(E;0,0)$ corresponds with
accuracy of order 10\% to the exact values of even levels already in the lowest
orders of approximation $n=1$ and $n=2$, i.e. when the path integral is replaced
by a line or double integral. The weights of the peaks approximate the values of
the squared modulus of the wave functions at $x=0$ with the same accuracy. The
accuracy depends on the type of the paths used. \end{abstract} \end{titlepage}

\section{Introduction}

Continual (functional) integrals,  or path integrals, introduced to quantum
mechanics by R.Feyn\-man \cite{1,2}, are widely used for constructing series in
perturbation theory, for finding quasi-classical  asymptotes, for quantizing gauge
fields \cite{3}--\cite{7}. Mathematical questions of the functional integral
theory are discussed in \cite{8,9}. Path integrals combined with Monte-Carlo
technique are being used for solving problems of quark confinement
\cite{10}-\cite{12}.

However, the question of applying path integral approximations to problems of practical
computations of quantum system spectra remains virtually unexplored.
First, not very successful, attempts of this kind are described by Kac\cite{13}.
An example of computation for a system of two particles sitting in a potential well
is considered in \cite{14}. Among more recent papers it is worth noting the work \cite{15}, where 
an estimate for the ground level from below is obtained with a help of path integral. Papers \cite{13}-\cite{15} employ ``Wick rotation'' and they allow obtaining information only about the ground level, moreover, for some examples
considered in \cite{13} the results disagree with exact values by a factor more than two. It seems that path integrals were applied for global spectrum analysis
in quasi-classical limit only \cite{16}-\cite{18}. It remains unclear if approximations to path integrals can have a practical value for calculations of
the Schr\"odinger equation spectra. To clarify the situation, it seems appropriate
to begin with simplest quantum systems and to investigate the influence of the path
types and of the order of approximations on the results.

Here we suggest the following approach to finding the energy levels
$E_j$ of a quantum system with the help of path integrals. The kernel of evolution
operator $\mathcal{K}(x_{b},T;x_{a},0)$ (i.e. the time-dependent Green function)
can be expressed via eigenfunctions $\Psi_j(x)$ of the Schr\"odinger equation:
\begin{equation}\label{eq1}
  \mathcal{K}(x_{b},T;x_{a},0)=\sum_j \psi_j(x_b)\psi_j^*(x_a)\exp(-iE_jT/\hbar).
\end{equation}
 For
$E$ shifted for an infinitesimal step to the upper half of the complex plane
and for $x_a=x_b=x_0$, equation (1) implies:
\begin{equation}\label{eq2}
  \Re\left\{
G(E,x_0)\right\}\equiv
\Re\left\{\int\limits_0^\infty\mathcal{K}(x_0,T;x_0,0)\exp(iET/\hbar)dT\right\}=
  \sum_J|\psi_j(x_0)|^2\pi\delta \left(\frac{E-E_j}{\hbar}\right),
\end{equation}
where $G(E,x_0)\equiv G(E,x_0,x_0)$ is the Green function of a stationary
Schr\"odinger equation.

On the other hand, the kernel (1) can be expressed via the path integral in configuration space \cite{1,2}:
\begin{equation}\label{eq3}
\mathcal{K}(x_{b},T;x_{a},0)= \lim_{n\to\infty}\mathcal{K}_n(x_{b},T;x_{a},0);
\end{equation}
\begin{equation}\label{4}
  \mathcal{K}_n(x_b,T;x_a,0)=C_{nF}(iT)^{-(n+1)/2} \int\limits_{-\infty}^{+\infty}\prod_{j+1}^n
  dx_j\exp[(i/\hbar)\int\limits_0^T L(\dot{q}_F,q_F,t)dT].
\end{equation}
Here
\begin{equation}\label{eq5}
  q_F(t)=q_*(t)+x_F(t),
\end{equation}
with $q_*(t)$ -- a segment of a straight line connecting $x_a$ and $x_b$, while
$x_F(t)$ is a broken line with vertices in
$x_j \; (j=1,2,\ldots n)$ separated by equal time intervals $T/(n+1)$;
$L(\dot{q}_F,q_F,t)$ is a Lagrangian, and
\begin{equation}\label{eq6}
  C_{nF}=[m(n+1)/(2\pi\hbar)]^{(n+1)/2}.
\end{equation}
Subscript $F$ in $q_F$ and $C_{nF}$ denotes the  type of the path $x_F(t)$ (Feynman).

Let us define a function $G_n(E,x_0)$, an approximation of the
$n$-th order to the exact Green function $G(E,x_0)$:
\begin{equation}\label{eq7}
  G_n(E,x_0)=\int\limits_0^\infty \mathcal{K}_n(x_0,T;x_0,0)\exp(iET/\hbar)dT.
\end{equation}

For $n\to \infty $ the real part of $G_n(E,x_0)$, i.e.  $\Re \{G_n(E,x_0)\}$, tends
to a sum of  $\delta$-functions in the r.h.s. of (2), according to (3). For finite
$n$ (for a sufficiently good approximation) the real part $\Re \{G_n(E,x_0)\}$
must have maxima near exact values of  $E_j$.

We put forward a technique for computations of $G_n(E,x_0)$ and carry out them
for harmonic and anharmonic oscillators in one- and multi-dimensional cases and
in a one-dimensional potential well of a shape $U_0\cosh^{-2}(q/a)$. We  find a number of analytical expressions  for $\Re \{G_1(E,x_0)\}$.

An unexpected result of our numerical experiments is the fact that a ``good''
approximation for finding the values of energy levels (with accuracy of order 10 per cent) is $G_n(E,x_0)$ already for $n=1$ and  $n=2$.

\section{Method of calculation}

The approximations 
$\mathcal{K}_n$ in (4) can be built based on an arbitrary complete set of functions, not necessarily broken lines. For example, using an expansion
over an  orthogonal set of sines one can replace the paths 
$q_F(t)$ in (4)  by
$$
q_s(t)=q_*(t)+\sum_{j=1}^n x_j\sin(\pi jt/T),
\eqno(5s)
$$
and the factor $C_{nF}$ by
$$
  C_{ns}=\frac{n!}{(2\pi)^{1/2}}\frac{\pi^{n/2}}{2^n}
          \left(\frac{m}{\hbar}\right)^{(n+1)/2}.
\eqno(6s)
$$
The factor 
$C_{ns}$ is determined from the condition that an approximate expression for the kernel  coincide
$\mathcal{K}_n$ with the exact $\mathcal{K}$ for all values of 
$n$ in case of a free particle.

For a Lagrangian
\begin{equation}\label{eq8}
  L(\dot{q},q,t)=m\dot{q}^2/2-U_0\varphi(q/a),
\end{equation}
with $U_0$ and $a$ being units of energy and length, we get from (4) and (5):
\begin{eqnarray}\label{eq9}
\lefteqn{
G_{n\ell}(E,x_0)=a^n C_{n\ell}\int\limits_0^\infty dT(iT)^{-(n+1)/2} \times } \nonumber \\
& &  \left[\int\limits_{-\infty}^{\infty}\prod_{j=1}^n dx_j
 \exp\left\{\frac{i}{\hbar}\left[\frac{\beta_\ell\sigma_{n\ell}}{T}
 - U_0Tf_\ell(x_0,x_1,\ldots,x_n)+ET\right]\right\}\right].
\end{eqnarray}
Here the subscripts  $\ell = F ,\; s$ correspond to paths $q_F(t)$ and $q_s(t)$
respectively. 
All distances  $x_j$ and $q_\ell$ in the Eq.(9) and herafter are measured in units 
$a$. The following notation is used: 
$$
\sigma_{nF}=\sum_{j=0}^n(x_{j+1}-x_j)^2, \qquad x_{n+1}=x_0, \qquad \beta_F=\frac{ma^2(n+1)}{2};
\eqno(10)
$$
$$
\sigma_{ns}=\sum_{j=1}^nj^2x^2_j, \qquad \beta_s=\pi^2ma^2/4;
\eqno(11)
$$
$$
f_\ell(x_0,x_1,...,x_n)=\int\limits_0^1\varphi\left(q_\ell(\tau)\right)dr, \qquad \tau=t/T.
\eqno(12)
$$

Integral over  $T$ in (9) can be done analytically interchanging the order of
integration over  $T$ and  $x_j$. For 
$E$ shifted to the upper semi-plane we have, according to [19],
$$
  G_{n\ell}(E',x_0) = B_{n\ell}\int\limits_{-\infty}^{+\infty} \prod_{j=1}^ndx_j(z/\sigma_{n\ell}^{(n-1)/2} H_{(n-1)/2}^{(1)}(z).
  \eqno(13)
$$
Here 
$B_{n\ell}=\pi a^n C_{n\ell}$; 
$H_{(n-1)/2}^{(1)}(z)$ is the Hankel function of the first kind;
$$
z=\gamma_\ell\{\sigma_{n\ell}[E'-F_\ell(x_0,x_1,...,x_n)]\}^{1/2},
\eqno(14)
$$
where 
$E'=E/U_0$, 
$\gamma_l=(2/\hbar)(\beta_\ell U_0)^{1/2}$.

The expression (13) is valid also as an approximation of the  $n$-th order of a
general Green function  $G(E; x_a,x_b)$. Then  $x_0=x_a, \quad x_{n+1}=x_b$ in
(10), and functions  $f_\ell$ in (12) and (14) depend both on  $x_a$ and 
$x_b$.

A useful tool for searching the levels can also be a spectral function,
$$
 Y(T)=\int\limits_{-\infty}^{+\infty}\mathcal{K}(x,T;x,0)dx=\sum_jg_j\exp
\left(-\frac{iE_jT}{\hbar}\right),
\eqno(15)
$$
where $g_j$ is the degeneracy of the $j$-th level.
Defining
$$
F(E)=\int\limits_0^\infty Y(T)\exp(iET/\hbar)dT,
\eqno(16)
$$
we see that  $F_{n\ell}(E)$, i.e. the  $n$-th order approximation to  $F(E)$ over
paths of type  $\ell$ is given by the expression
$$
F_{n\ell}(E)=aB_{n\ell}
\int\limits_{-\infty}^{+\infty}\prod_{j=0}^ndx_j(z/\sigma_{n\ell})^{(n-1)/2} H^{(1)}_{(n-1/2}(z).
\eqno(17)
$$
This expression differs from (13) only by raising the order of integration by
unity. Maxima of the real part of  $F_{n\ell}(E)$ must lie near the exact energy
levels while the height of the maxima must reflect the degree of the level
degeneracy.

The real part of the integral in (13) can be written in the form
$$
R_{n\ell}(E')\equiv \Re\{G_{n\ell}(E',x_0)/B_{n\ell}\}=\int\limits_D\prod_{j+1}^n dx_j(z/\sigma_{n\ell})^
{(n-1)/2}J_{(n-1)/2}(z),
\eqno(18)
$$
where  $J_{(n-1)/2}(z)$ is Bessel function. The domain  $D$ of integration over 
$x_j$ in (18) is determined by the condition
$$
D: \qquad f_\ell(x_0,x_1,...x_n)\leq E'.
\eqno(19)
$$
If the inequality (19) is violated then the integrand in (18) is purely imaginary.

For even  $n=2k+2 \quad (k=0,1,2\ldots)$   the Bessel function $J_{(n-1)/2}(z)$
reduces to elementary functions and we have from (18):
$$
R_{n\ell}(E')=(2/\pi)^{1/2}\int\limits_D\prod_{j=1}^ndx_j\sigma_{n\ell}^{(1-n)/2}
 z^{2k+1}
\left(-\frac{1}{z}\frac{d}{dz}\right)^k \frac{\sin z}{z}, \quad n=2k+2.
\eqno(20)
$$

The expression for real part of $F_{n\ell}(E)$ is similar to (18) and (20) with
$\prod_{j=1}^n$ replaced by  $\prod_{j=0}^n$

The formulae (9 -- 18) can be easily generalized for a multi-dimensional case. If
the dimension of vectors  $\vec x$ and  $\vec q$ is  $\nu$ then
$$
G_{n\ell\nu}(E,\vec{x}_0)=a^{\nu n}C^\nu_{n\ell}\int\limits_0^\infty dt (iT)^{-\nu(n+1)/2}\times
$$
$$
\left[\int\limits _{-\infty}^{+\infty}\prod_{j=1}^n d\vec{x}_j\exp\{(i/\hbar)[\beta_\ell
\sigma_{n\ell\nu/}/T-U_0 T f_{\ell\nu}(\vec{x}_o,\vec{x}_1,...,\vec{x}_n)+ET]\}\right];
\eqno(9a)
$$
$$
\sigma_{nF\nu}=\sum_{j=0}^n(\vec{x}_{j+1}-\vec{x}_j)^2;
\eqno(10a)
$$
$$ \sigma_{ns\nu}=\sum_{j=0}^n j^2
\vec{x}^2_j;
\eqno(11a)
$$
$$
f_{\ell\nu}(\vec{x}_0,\vec{x}_1,...\vec{x}_n)=\int\limits^1_0 \varphi( \vec{q}_\ell(\tau))dr;
\eqno(12a)
$$
$$
G_{n\ell\nu}(E')=B_{n\ell\nu}\int\limits _{-\infty}^{+\infty}\prod_{j=1}^n d\vec{x}_j(z_{\nu}/\sigma_
{n\ell\nu})^{\frac{\nu(n+1)}{2}-1} \times H^{(1)}_{ \frac{\nu(n+1)}{2}-1} (z_\nu);
$$
$$
B_{n\ell\nu}=\pi a^{\nu n}C^\nu_{n\ell} (2\beta_\ell/\hbar)^{1- \frac{\nu (n+1)}{2}};
$$
$$
z_\nu =\gamma_\ell\{\sigma_{n\ell\nu}[E'-f_{\ell\nu}( \vec{x}_0,...\vec{x}_n)]\}^{1/2};
\eqno(14a)
$$
$$
R_{n\ell\nu}(E')=\int\limits_D \prod_{j=1}^n d\vec{x}_j(z_{\nu}/\sigma_{n\ell\nu})^{\frac{\nu(n+1)}{2}-1}
\times J_{ \frac{\nu(n+1)}{2}-1}(z_{\nu}).
\eqno(18a)
$$

In this work the calculations were done using formulae (18), (18a) and (20). We
omit the subscript  $\nu$ below when  $\nu=1$.

\section{Results of calculations}

\subsection{Harmonic oscillator}

The exact expression for the kernel 
$\mathcal{K}$ of one-dimensional harmonic oscillator is well-known, still it is interesting to consider the behavior of 
$R_{n\ell}(E)$ for low orders of approximation. 
For
$$
\varphi(q)=q^2, \quad U_0=\hbar \omega/2\pi
\eqno(21)
$$
it follows from (18) that, if $n=1$,  
$$
R_{1\ell}(E')=A_\ell\xi^{1/2}_\ell J_{1/4}(\xi_\ell)J_{-1/4}(\xi_\ell),
\eqno(22)
$$
where
$$
A_F=\pi/\sqrt{2}, \qquad \xi_F=\sqrt{3}E'/2\pi;
$$
$$
A_s=\sqrt{2\pi},   \qquad         \xi_s=E'/4.
$$

In case of evenly symmetric potentials,  $U(q)=U(-q)$,  for odd levels, the
wavefunctions vanish at  $q=0$, $\psi_j(0)=0$, therefore, according to (2) maxima
of the r.h.s. of expression (22)  must correspond only to even energy levels 
$E'_j$ for  $j=0,2,4, \ldots$. Those exact values are shown by arrows in Fig.1
where the dependence of  $R_{1F}$ and  $R_{1s}$ on  $E'$ is plotted according to
(22).

\FIG f1 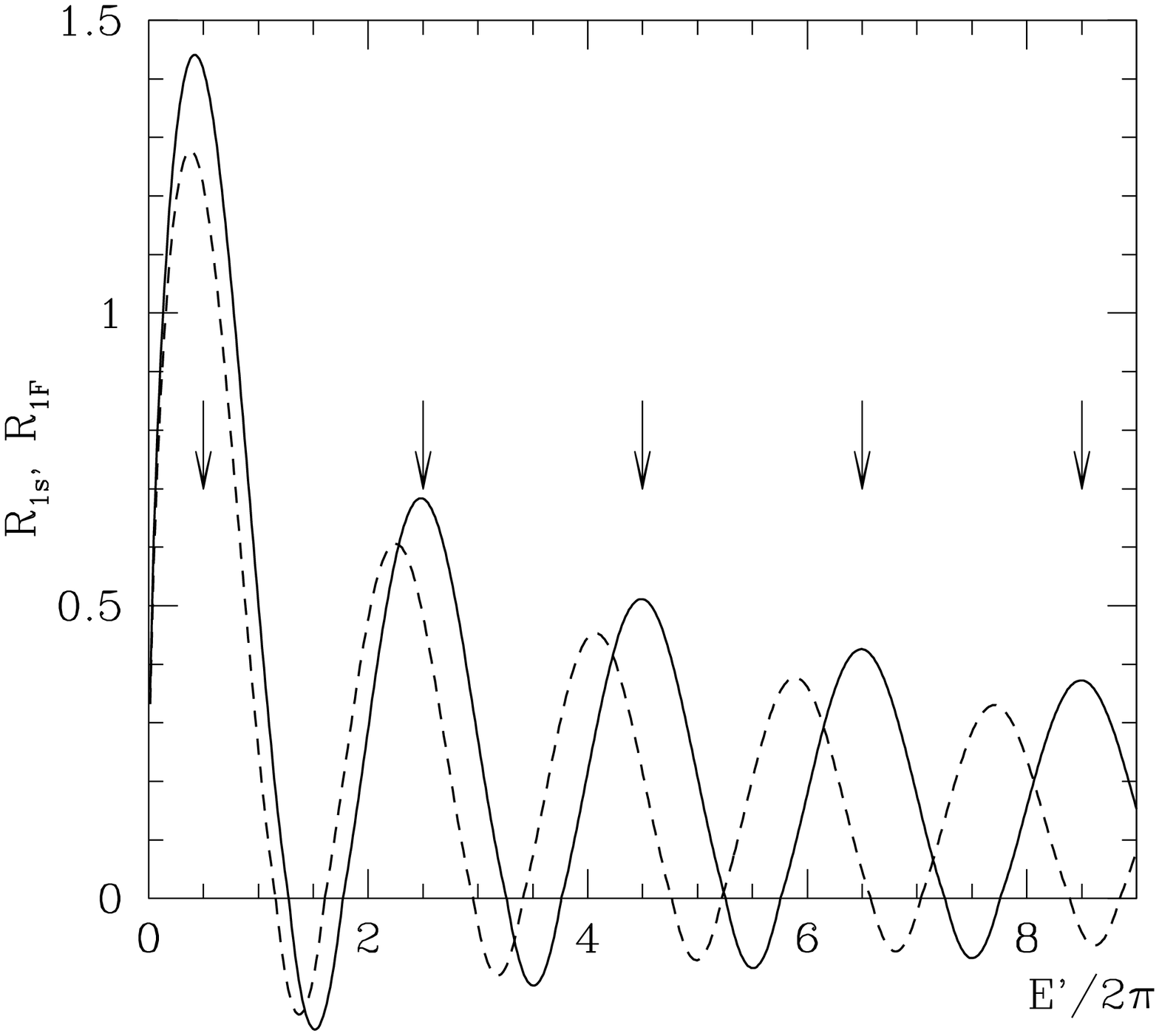 [width=10cm] Dependencies $R_{1\ell}(E')$ for
the one-dimensional harmonic oscillator (21). Solid 
$\ell=s$, dashed $\ell=F$. The exact energy values of even levels are shown by arrows.

One can see that the dependence 
$R_{1\ell}(E')$ has clear peaks located in correspondence with the levels of the oscillator. E.g., for expansion over sines (5s)  the location of the first maximum
deviates from the exact value of the ground level by less than 15\%.

The behavior of $R_{n\ell}(E')$ (which is proportional to $\Re\{G_{n\ell}(E')\}$ )
must reflect the properties of a sum of 
$\delta$-functions. The location of a 
$\delta$-function is determined by an integral relation. Therefore it is more natural to expect that a level corresponds not to a maximum of  $R_{n\ell}(E')$,
but to an average of the integral (i.e. a median value) of  $R_{n\ell}(E')$ between
two minima of the dependence  $R_{n\ell}(E')$. We call this median value 
a median of the peak, and the area under the curve  $\Re\{G_{n\ell}(E')\}$ between two adjacent minima of  $R_{n\ell}(E')$  will be called a weight of the peak.

The median of the ground level peak calculated from (22) is equal to $3.08$
which differs form the exact value,  $\pi$, by 2\%.

For
$E'\gg 1$ we have from (22):
$$
R_{1\ell}(E')\approx A_\ell\pi^{-1}\xi_\ell^{-1/2}[2^{-1/2}+\cos(2\xi_\ell-\pi/2)],
\eqno(23)
$$
i.e. maxima and medians of the peaks  
$R_{1F}(E')$ are at 
$E'_j=(\pi^2/\sqrt{3})(j+1/2))$, and of 
$R_{1s}(E')$ at 
$E'_j=2\pi(j+1/2)$, for large even 
$j$. Thus, medians of peaks of 
$R_{1s}(E')$ coincide with exact values of oscillator levels for 
$E'\gg 1$, while the expansion over broken lines gives a spectrum which is qualitatively correct, but with a relative error 
$\sim 10$\% (approximately the same as for the ground level).

With account of the coefficient 
$B_{1s}$ in (13), we obtain from (22) for 
$\ell=s$ that the  weight of the peak of the ground level is equal to 
$1.776(\hbar m \omega)^{1/2}$. This differs from the exact value, 
$\pi\hbar |\psi_0(0)|^2=(\pi \hbar m \omega)^{1/2}$, by less than 1\%.

In general case:
$$
|\psi_{2k}(0)|^2=\frac{(2k)!}{\sqrt \pi 2^{2k}(k!)^2} \frac{m\omega}{\hbar},
$$
therefore, we find an asymptotic formula 
$\pi\hbar|\psi_{2k}(0)|^2=(\pi\hbar m\omega/k)^{1/2}$ 
for levels with large number $j=2k$. According to (23), for 
 $\ell=s$ the weights of peaks coincide with this exact value.  For  $\ell=F$ the
difference between the peak weight and the exact value 
$\pi\hbar|\psi_{2k}(0)|^2$ is of order  $\sim 10$\%.

Now let us turn to approximations of the second order ($n=2$). The results of
numerical calculations of  $R_{2F}(E')$ and $R_{2s}(E')$ based on the formula (20)
are shown in Fig.2. 
For  $n=1$ the graphs of $R_{1F}(E')$ and $R_{1s}(E')$  were
similar to each other, but now we see a drastic difference in the behavior
of $R_{2\ell}(E')$. The
function $R_{2F}(E')$, built with the use of broken paths, has ``false'' maxima,
while $R_{2s}(E')$  demonstrates quite a regular behavior showing a tendency to a
sum of  $\delta$-functions.

\FIGG f2 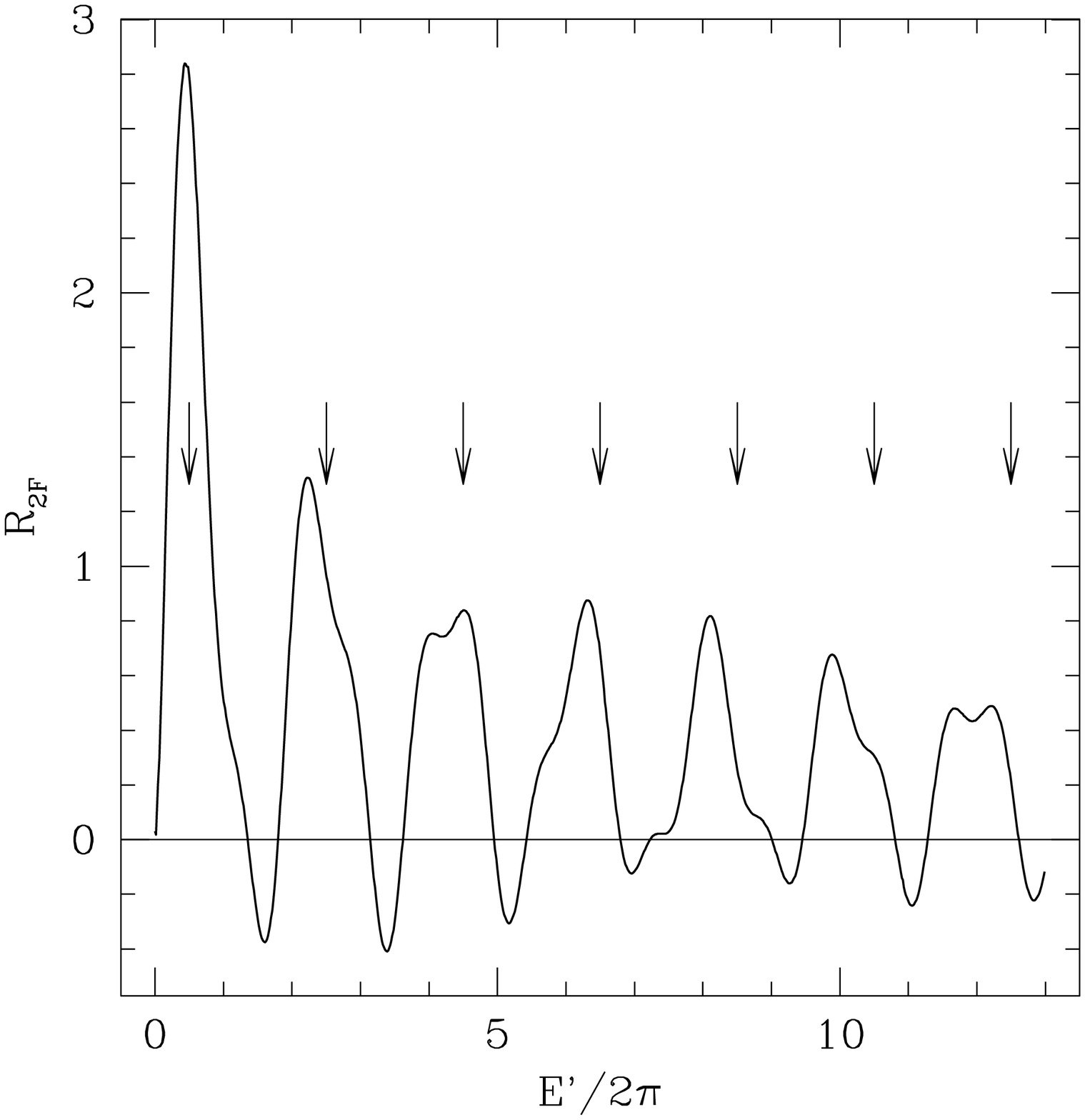 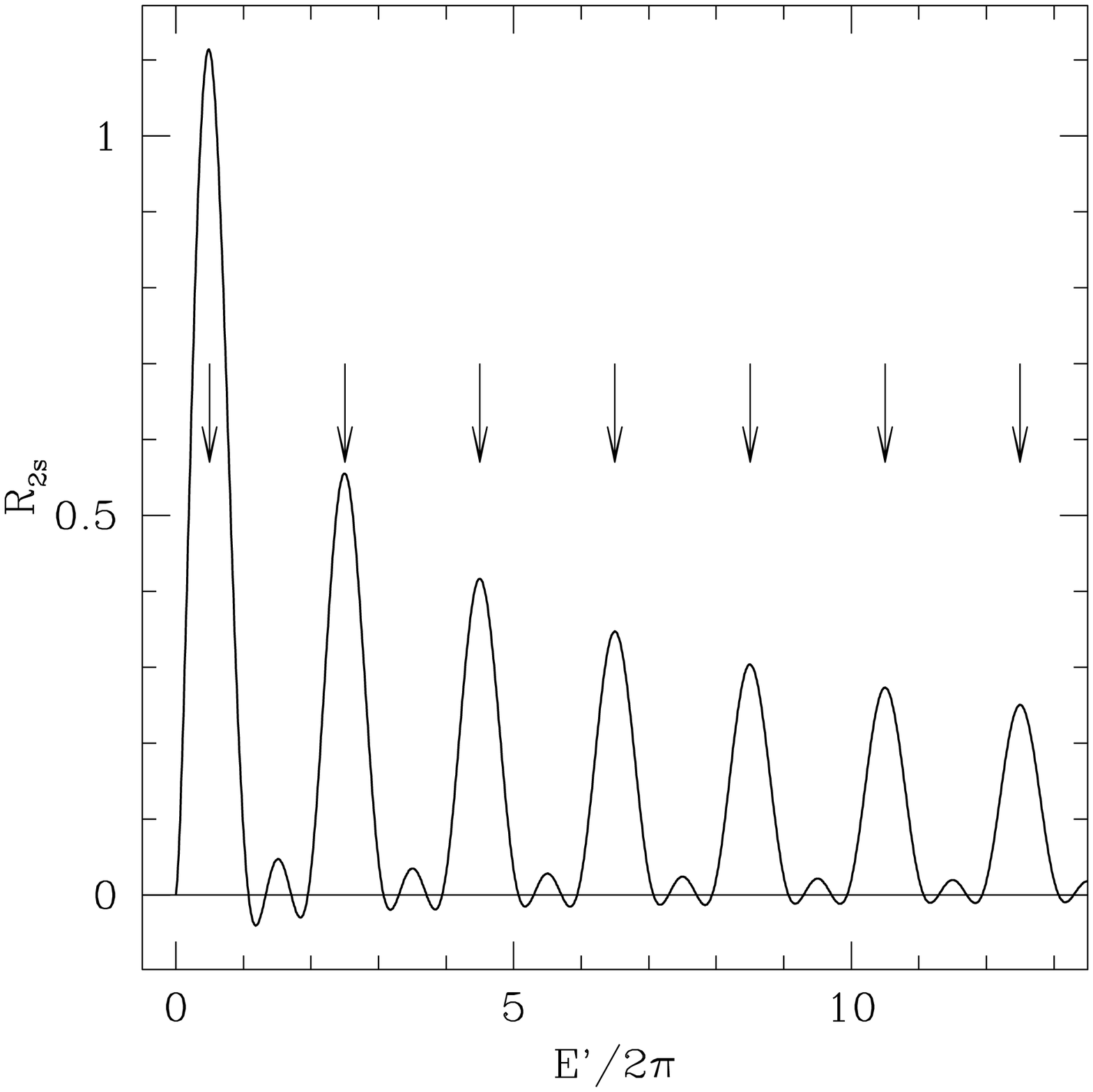 [width=.5\textwidth] Dependencies $R_{2\ell}(E')$ for the  harmonic oscillator potential (21). Left: for $\ell=F$;
right: for $\ell=s$. The exact energy values of even levels are shown by arrows.

It is likely that good results obtained with the paths (5s) for 
$n=2$ are due to the fact that they are optimum for approximations of path integrals over the Wiener measure [20, 21].
For other potentials, e.g. for a box with reflecting 
walls, the behavior of the function  $R_{nF}(E')$ is quite regular (see below) and it has no ``false'' maxima.
Hence, the system of optimum functions should depend on the shape of the potential
well.

\subsection{Anharmonic oscillators}

For  power-law potentials 
$$
\varphi (q)=q^N
\eqno(24)
$$
and even $N$ we present the results of computations  of $R_{n\ell}(E')$ for 
$n=1, \; 2$. Instead of $E'$, here it is convenient to introduce a variable 
$$
y=E'^{(N+2)/2N}(\pi^2ma^2U_0/\hbar^2)^{1/2}I_N^{-1/N},
\eqno(25)
$$ 
where for $N=2p: \quad p=1, \; 2,\; 3 \ldots$ the expression for 
$I_N$ reads $I_N=(2p-1)!!/(2p)!!$. The graphs  $R_{2s}(y)$ for 
$N=4,\; 10,\; 50$ are shown in Fig.3.

  \begin{figure}[!h]
    \begin{center}
    \includegraphics*[width=.5\textwidth]{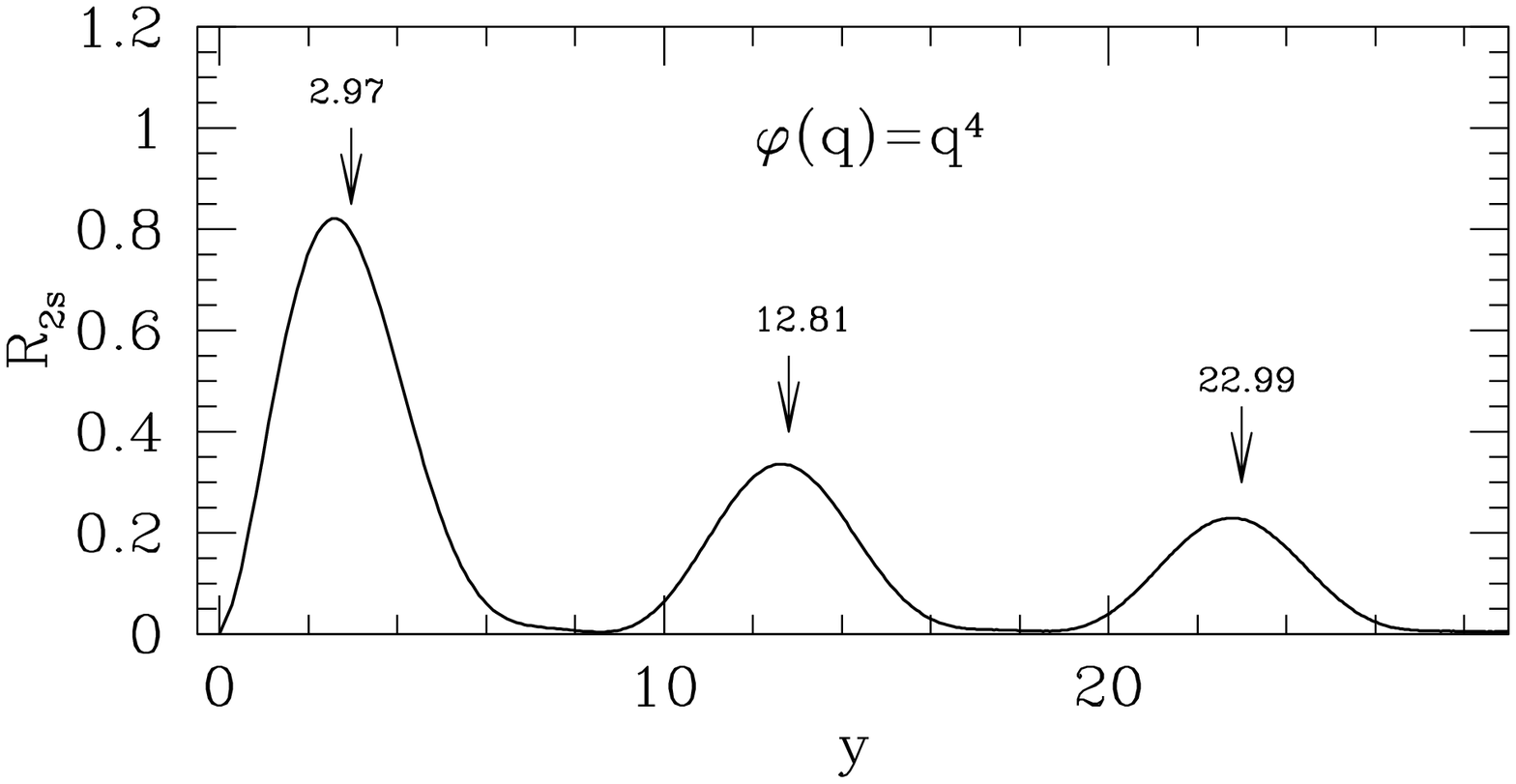} 
    \includegraphics*[width=.5\textwidth]{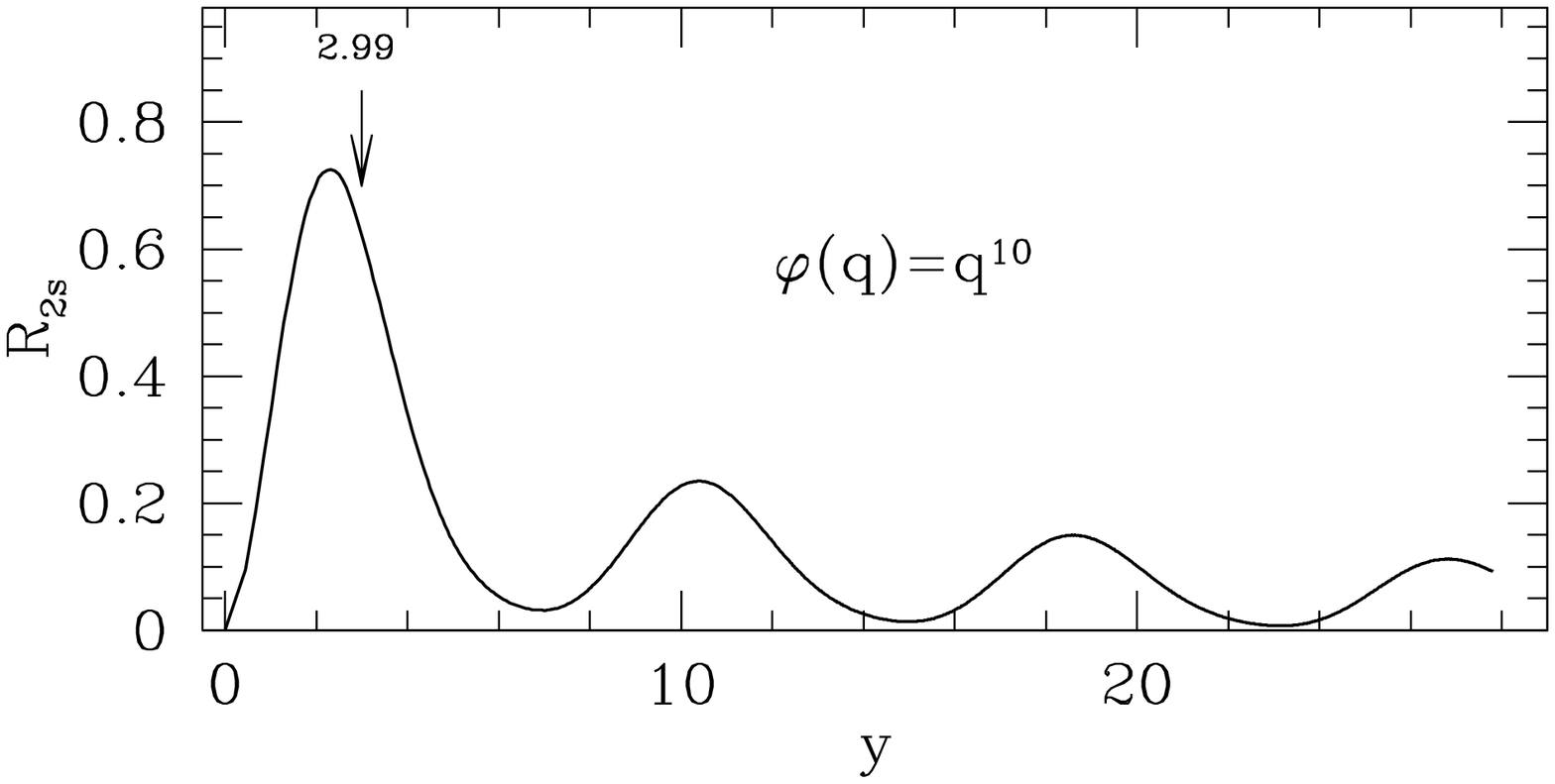} 
    \includegraphics*[width=.5\textwidth]{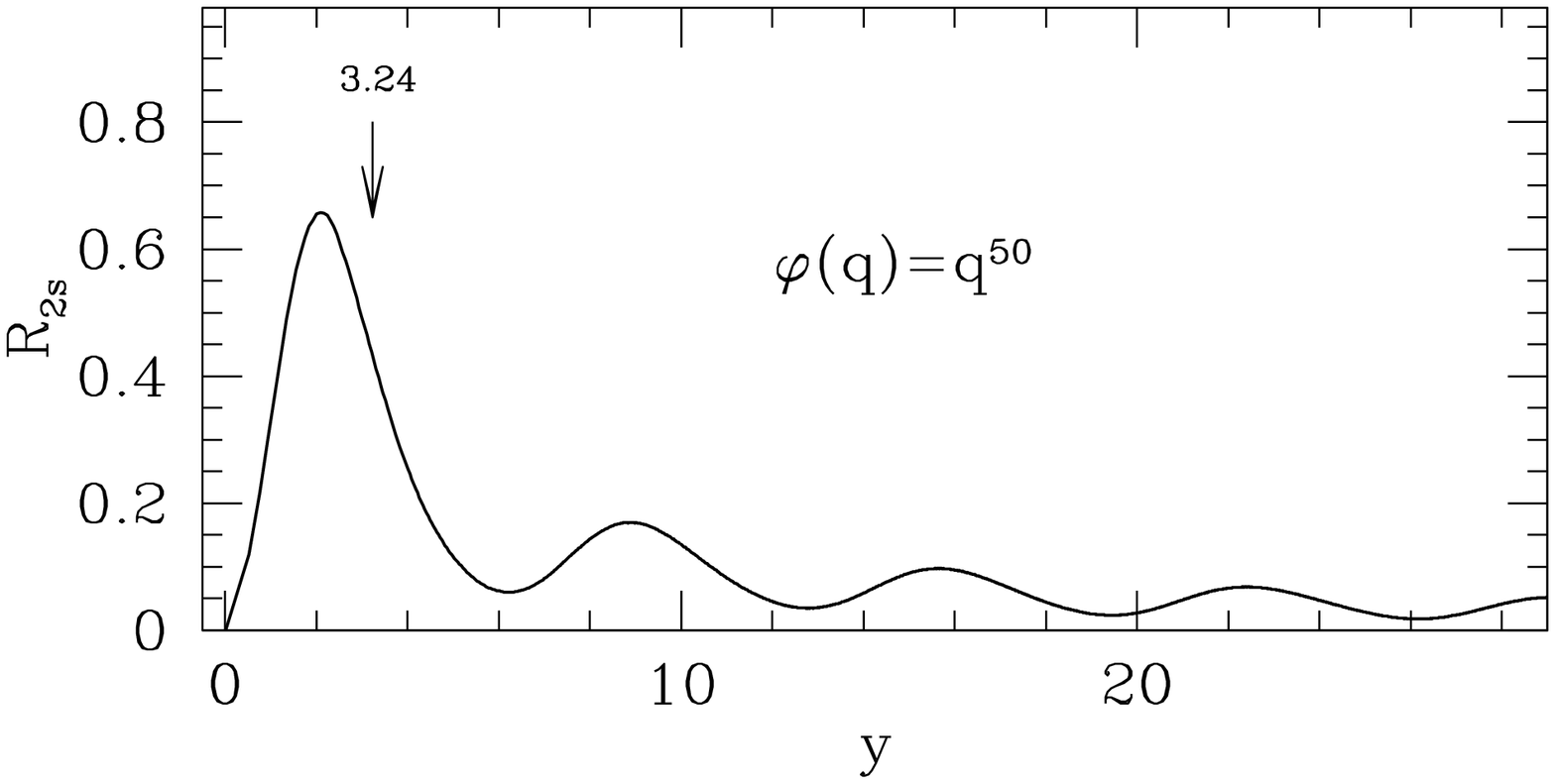}
    \end{center} 
    \caption{\label{f3} Anharmonic oscillators (24). Arrows point to exact values
  of even levels \cite{15},\cite{22}.}          
  \end{figure}

The case  $N=\infty$ (i.e. a rectangular well of infinite depth) allows for an
analytical expression for  $R_{1\ell}(E)$. It follows from (18) that
$$
R_{1\ell}(E)=2\eta^{-1}_\ell\int\limits_0^{\eta_\ell}J_0(x)dx,
\eqno(26)
$$
where 
$\eta_F = 2\sqrt{2}\,y/\pi$, 
$\eta_s=y=\pi(E m)^{1/2}a/\hbar$, and 
$2a$ is the width of the well.

The integral (26) can be expressed through the Struve functions, yet it is more convenient to use tables and expansions for this integral given, e.g., in [23].

For $\eta_\ell \gg 1$ , that is for $E \gg E_0$, expression (26) implies
$$
R_{1\ell}(E)\approx 2\eta_\ell^{-1}[1+(2/\pi \eta_\ell)^{1/2}\cos(\eta_\ell-3\pi/4)].
\eqno(27)
$$

Maxima (and medians) of peaks of  $R_{1F}(E)$ are reached at  
$E_{2k}^{1/2}=\hbar(2k+3/4)/\sqrt{8ma^2}$, whereas the exact values of energy levels number  $2k$ are
$E_{2k}^{1/2}=\hbar (2k+1)/\sqrt{8ma^2}$, for $k=0, \; 1, \; 2 \ldots$.
The results of numerical calculations for the peak medians 
$y_{1F}$ and $y_{1s}$, corresponding to the ground level for oscillators with various values of 
$N$, are given in Table 1 for 
$n=1, \; 2$.

The data in Table 1 show that the difference between approximate and exact values for the ground level increases with growing $N$ for 
$\ell=s$ and decreases for $\ell=F$.

Medians of the peaks for higher levels also have a systematic shift relative to
the exact values.  The relative error for high levels grows from 0\% for $N=2$ up to about 10\% for $N \to \infty$ when $\ell=s$ and $n=1$; it falls from $\sim$10\% down to zero for $N$ changing from 2 up to $\infty$ when $\ell=F$ ($n=1$).

In the latter case,  $\ell=F$, for  $N=\infty$ the weight of the first peak is
7.4\% less than the exact value $\pi\hbar/a$, and for high levels the weight tends
to the exact value $\pi\hbar|\psi_{2k}(0)|^2 = \pi\hbar/a$.

Some deterioration of the results for  $N=\infty$ in comparison with the case 
$N=2$ can be probably explained as follows.

\begin{table}
 \begin{center}
 \caption{Peak medians in comparison with the exact values 
$y_{\rm exact}$ [15] normalized as in (25)}
 \label{tb:Xname}
  \begin{tabular}{|l|l|l|l|l|l|} \hline
   \multicolumn{1}{|c}{N} &
   \multicolumn{1}{|c}{2} & 
   \multicolumn{1}{|c}{4} &
   \multicolumn{1}{|c}{10} & 
   \multicolumn{1}{|c}{50} &
   \multicolumn{1}{|c|}{ $\infty$} \\ 
\hline \hline
   $y_{\rm exact}$ & 3.1416  & 2.9663 &  2.9899	& 3.2431 & 3.4894 \\
   $y_{1s}$        & 3.08    & 2.90   &  2.82	& 2.82   & 2.832 \\
   $y_{1F}$        & 2.79    & 2.75   &  2.84	& 3.02   & 3.145 \\
   $y_{2s}$        & 3.12    & 2.94   &  2.86	& 2.85   &  --- \\
\hline
  \end{tabular}
 \end{center}
\end{table}

For  $N=2$ the straight line  $q_*(t)=0$, which serves as a leading term of
expansion (5), is the unique classical path going from  $x_a=0$ to  $x_b=0$ in
time  $T$, if $T\ne 2\pi k/\omega$ ($k=0, \; 1, \; 2 \ldots$),  that provides for
the minimum of the action functional.  When  $N\ne 2$ (in particular for 
$N=\infty$) for any (even infinitesimal)  $T>0$ there exists infinite (countable)
set of classical paths (without a reflection, with one reflection, with two
reflections, etc.) [24]. The influence of additional classical paths seems to grow
for  $N\to\infty$.

\subsection{Influence of continuum spectrum}

Above are considered the examples of potentials for which there exists only a discrete spectrum of eigenvalues of energy. For continuum  spectrum the relation (2) is not applicable. It is therefore seems interesting to consider a well of finite depth, since this problem has both discrete and continuum ranges of a particle spectrum. 

In a well with the potential 
$$
\varphi(q)=-\cosh^{-2}(q/\gamma), 
\eqno(28)
$$
for distance unit $ a=\pi^{-1}\hbar(mU_0)^{-1/2}$ in Eq.(8),
already the first order of approximation using sinusoidal paths provides a reasonable accuracy. Plots of $R_{1s}(E')$ for various values of 
$\gamma$ are presented in Fig.4. 
One can observe the growth of the number of bound levels 
$E'_n<0$ with growing depth of the well (growing parameter $\gamma$),
$$
 E'_n=-\frac{\pi^2}{2\gamma^2}\left(n+\frac{1}{2}-\frac{1}{2}
 \sqrt{1+\frac{8\gamma^2}{\pi^2}}\right)^2 \; , 
\quad n\le -\frac{1}{2}+\frac{1}{2} \sqrt{1+\frac{8\gamma^2}{\pi^2}}.
$$

\begin{center}%
    \includegraphics*[width=10cm]{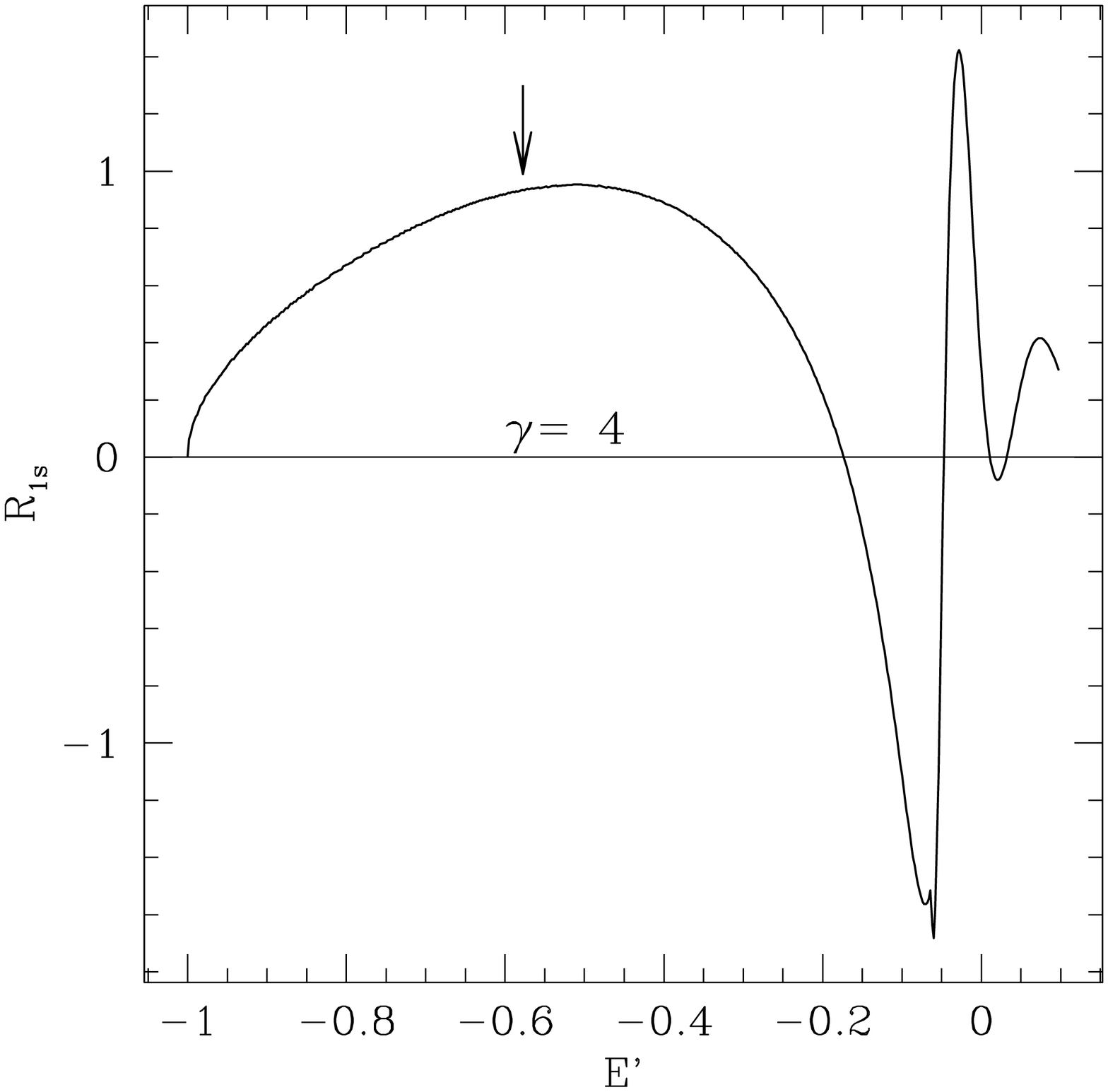} \\%
     Figure 4a:
  \end{center}
\begin{center}%
    \includegraphics*[width=10cm]{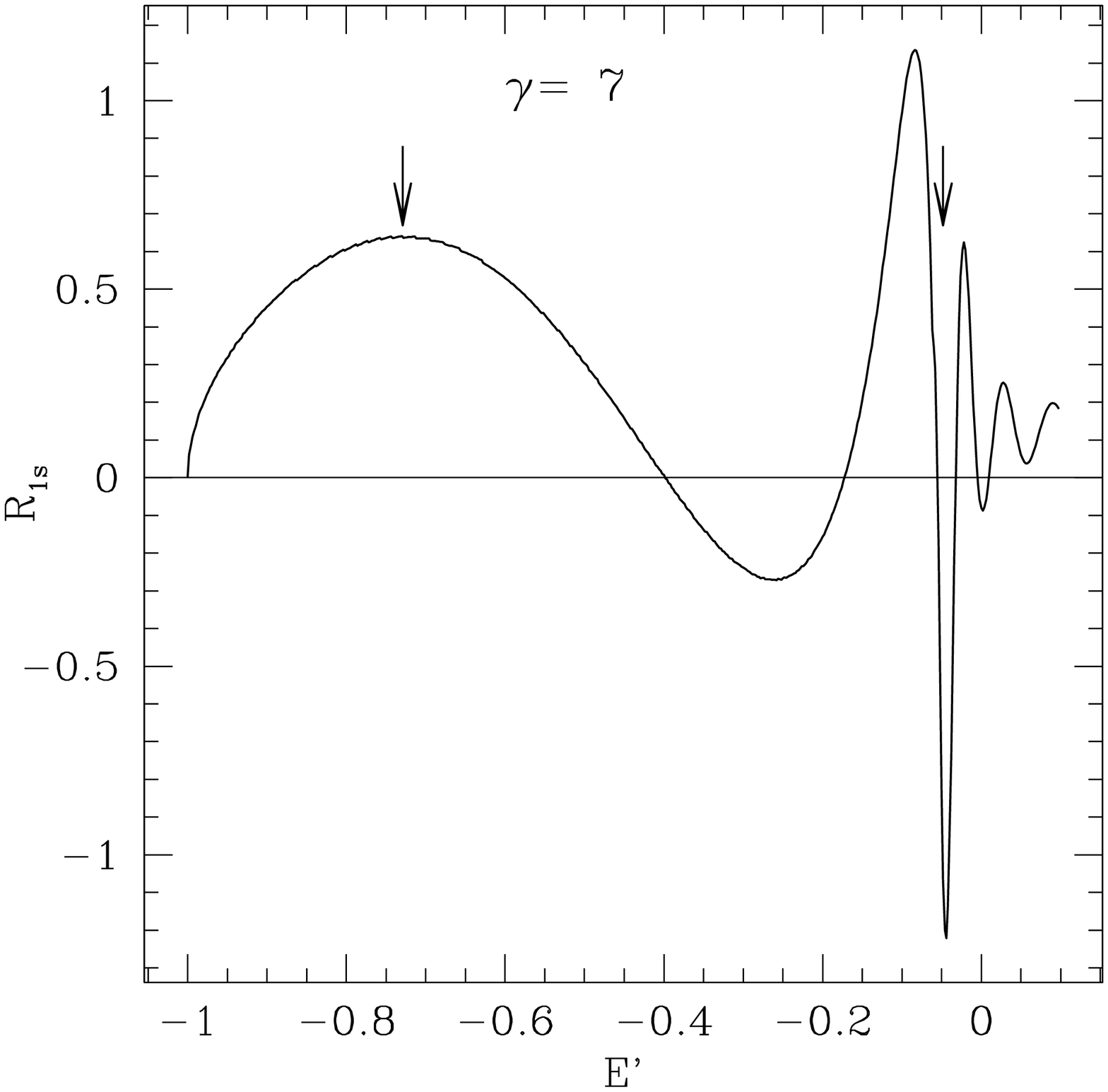} \\%
     Figure 4b.
  \end{center}
\begin{center}%
    \includegraphics*[width=10cm]{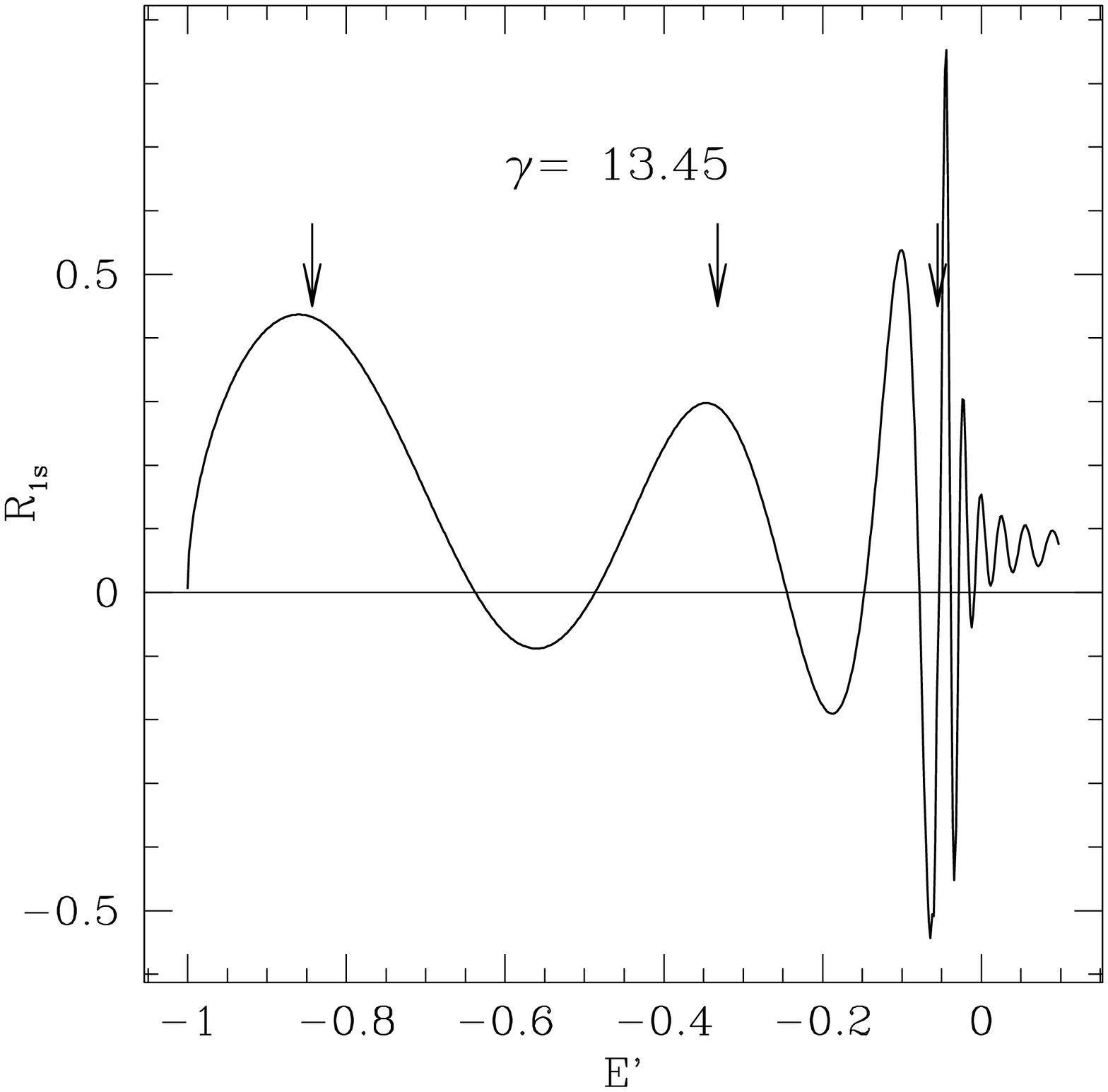} \\%
     Figure 4c.
  \end{center}
\begin{center}%
    \includegraphics*[width=10cm]{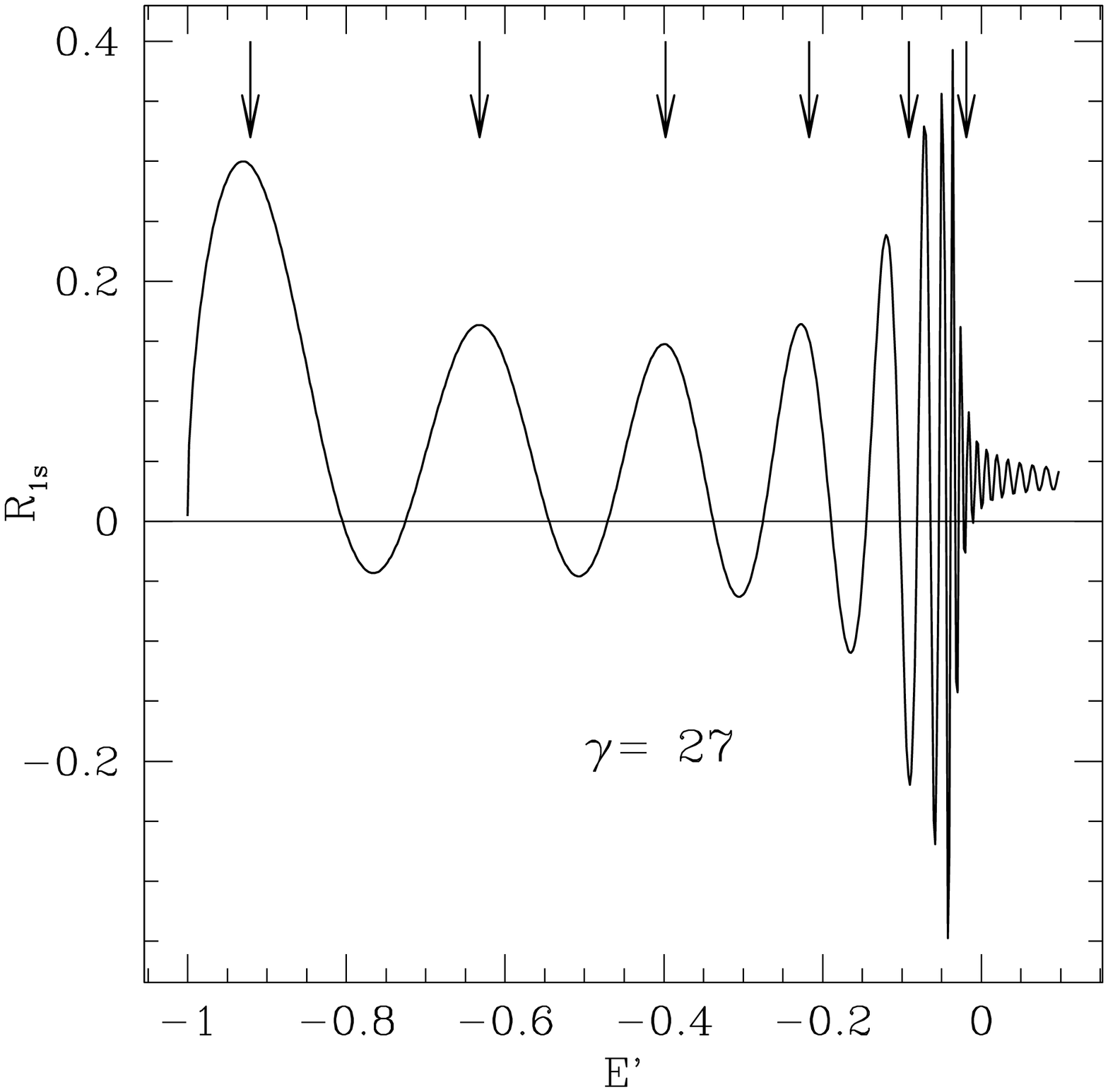} \\%
     Figure 4d.
  \end{center}

It is important that not only approximations for
ground levels are in good agreement with exact values [since for small 
$q$ the potential (28) tends to a harmonic oscillator potential], but there is also
a correct description of intermediate levels. The uppermost levels feel the influence of continuum. We get from (18) for 
$n=1$ an asymptote for 
$E' \to -0$: 
$$
R_{1s}(E')\sim \gamma^{-1}(-E')^{-1/2}\sin [\gamma\pi^{-1}(-E)^{-1/2}],
\eqno(29)
$$
i.e. there appear oscillations with growing frequency and amplitude in the
function $R(E')$ for $E'\to-0$. 
The square root singularity $R(E) \sim E^{-1/2} $ is typical for continuum 
spectrum, one can easily check this taking the integral in the left hand side of (2) for the evolution operator kernel of a free particle.
In practice this singularity can be easily removed putting the system under consideration into a wide potential well, which transforms the continuum into
a discrete spectrum.
However, the distortion of uppermost levels is inevitable in this case also.

\subsection{A case of higher dimensions}

For oscillators with a potential 
$$
\varphi (\vec{q})=\vec{q}\,^2 \; ,
$$
if the dimension of the problem is 
$\nu$, we obtain from (18a) the following relation:
$$
R_{1s\nu}(E')=\Omega_\nu \int\limits_D (\gamma_s E')^{(\nu-1)/2} \sqrt{2E'}
(1-x^2_1)^{(\nu-1)/2}  J_{\nu-1}\left(\gamma_s\sqrt{2}E'x_1(1-x^2_1)^{1/2}\right) dx_1
$$
$$
=\frac{\Omega_\nu \gamma_s^{\nu-1}}{\sqrt{2} }(E')^{\nu/2}\int\limits_0^\pi
\left(\cos\frac{\theta}{2}\right)^\nu J_{\nu-1}\left( \frac{\gamma_s E'}{\sqrt{2}}\sin\theta\right)d\theta,
\eqno(30)
$$
where 
$\Omega_\nu=S_\nu/r^{\nu-1}$ and 
$S_\nu$ is the area of the 
$\nu$-dimensional sphere.

Using the tables of integrals [25], we have from (30), leaving aside a constant
factor:
$$
R_{1s2}(E')\propto E' J^2_{1/2}(\xi);
$$

$$
R_{1s3}(E')\propto (E')^{3/2}\{3J_{3/4}(\xi) J_{5/4}(\xi)-J_{1/4}(\xi)J_{7/4}(\xi)\};
$$

$$
R_{1s4}(E')\propto (E')^2 \{3J^2_{3/2}(\xi)-J_{1/2}(\xi)J_{5/2}(\xi)\}.
\eqno(31)
$$
Here 
$\xi=\sqrt{2}\gamma_s E'$; 
$U_0=\hbar\omega/2\pi$.

For multi-dimensional wells with perfectly reflecting walls with the potentials
like 
$$
  \phi(\vec q)=0 \quad \mbox{for} \quad |\vec q| \le a
$$
and
$$
  \phi(\vec q)=+\infty \quad \mbox{for} \quad |\vec q| > a \; ,
$$
we have
$$
R_{1f\nu}\propto y^{\nu-2}\int\limits_0^{2\sqrt{2}y\pi} J_{\nu-1}(x)dx,
\eqno(32)
$$
where $y=2\sqrt{2}(ma^2 E)^{1/2}/\hbar$.

For 
$\nu=2$ and 
$\nu=3$ it follows from (32) that
$$
R_{1f2}\propto 1-J_0(2 \sqrt{2}y/\pi) \; ;
$$
and
$$
R_{1f3}\propto y\left[\int\limits _0^{2\sqrt{2}y\pi} J_0(x)dx-2J_1 (2\sqrt{2}y/\pi)\right].
\eqno(33)
$$

The formulae (31) and (33) imply that medians of the peaks for ground levels
are close to exact values of energy.
The energy dependences of the peak weights obtained with these formulae
also reproduce correctly the behavior of the squared module of the wavefunction:
we have for oscillators 
$|\psi_{2k}(0)|^2\propto E_{2k}^{(\nu-2)/2}$, and for wells,
$|\psi_{2k}(0)|^2\propto E_{2k}^{(\nu-1)/2}$.

\section{Conclusions}

The numerical experiments and analytical examples considered in this work allow us to draw some conclusions.

\begin{enumerate}
\item  Replacing a path integral by its approximation with an integral of 
finite dimensionality may be an effective means for computing spectra
of multi-dimensional quantum systems. 
It is remarkable that qualitatively, and even quantitatively 
(with accuracy 
$\sim 10$\%), correct information 
on the energy levels and on values of 
$|\psi|^2$  can be obtained already with approximations of low order, 
$n=1, \; 2$, that is with a small amount of computing.
\item Our preliminary computations, using Korobov numerical quadrature for 
$n$ up to $n=16$, show that the accuracy of method grows not faster than 
$n^{-1/2}$.
\item It would be interesting to generalize the method for the case of
many-body systems with account of particle statistics, probably with the help
of Monte-Carlo methods, and also to apply this technique for computations
of thermodynamic functions of non-ideal systems.
\end{enumerate}



\section*{Addendum}

{\small

The text above is almost a literal translation of a Russian ITEP preprint
111-82. Only a few misprints are corrected and a well known formula
for $E_n$ in modified P\"oschl-Teller potential  [a] is added after (28).

During the two decades, that passed after the preprint was circulated,
there appeared many papers on the subject of numerical applications of path
integrals. Yet it seems that our results are still interesting because the
technique of numerical analysis that we have undertaken, namely, using  our formulae (18) and (19), was not explored by other workers for obtaining information
on excited levels. Moreover, we found that the lowest orders in (18) and (19) 
are sufficient for obtaining reasonable results. Thus, this direction of research
is still promising. 

Some historical references, missing in our text, are added below [a-e], as well
as references to relevant papers [f-v] published after this report was issued in 1982.

We thank Andrei Smilga for discussions that stimulated the translation  of this paper into English and its posting in arXiv. S.B. is very grateful to H.Takabe
for his kind hospitality at ILE, Osaka, where a major part of this version
of the paper has been prepared, and to V.Zhakhovskii for useful hints on recent literature.

\subsection*{ Additional References}

\renewcommand{\labelenumi}{\alph{enumi}.}
\begin{enumerate}
\item P\"oschl G., Teller E. - Bemerkungen zur Quantenmechanik des anharmonischen Oszillators. - Zs.Phys., 1933, v.83, p.143.
\item Davison B. - On Feynmann's `integral over all paths'.
	- Proc. R. Soc. Lond. A, 1954, v. 225, p.252.
\item Gelfand I.M., Jaglom A.M. -  
	Integration in Functional Spaces and its
	Applications in Quantum Physics. - J.Math.Phys., 1960, v.1, p.48.
\item Nelson E. - Feynman Integrals and the Schr\"odinger Equation. - 
            J.Math.Phys., 1964, v.5,  p.332.
\item Duru I.H., Kleinert H. - Solution of the Path Integral for the H-Atom. -
	Phys.Lett. B, 1979,  v.84, p.185.
\item Ho  R., Inomata A. - 
        Exact-Path-Integral Treatment of the Hydrogen Atom. - Phys. Rev. Lett.  1982, v.48, p.231.
\item De Raedt H., De Raedt B. - Applications of generalized Trotter Formula.
	- Phys. Rev. A, 1983, v.28, p.3575.
\item Takahashi M., Imada M. - Monte Carlo Calculation of Quantum Systems. -
	J. Phys. Soc. Japan, 1984, v.53,  p.963. 
\item Takahashi M., Imada M. - Monte Carlo Calculation of Quantum Systems. 
        II. Higher Order Correction. -
	J. Phys. Soc. Japan, 1984, v.53,  p.3765. 
\item Doll J.D., Coalson R.D., Freeman D.L. - Fourier Path-Integral Monte Carlo
	Methods: Partial Averaging. -  Phys. Rev. Lett., 1985, v.55, p.1.
\item Filinov V.S. -  Calculation of the Feynman integrals by means of the Monte Carlo method. - Nucl. Phys. B, 1986, v.271, p.717. 
\item Kleinert H. - How to do the Time Sliced Path Integral for the H Atom. -
	Phys.Lett. A, 1987, v.120, p.361.
\item Grosche C. - Path Integral Solution of a Class of Potentials Related to the
	P\"oschl-Teller Potential. - J.Phys.A: Math.Gen.,  1989, v.22, p.5073.
\item Grosche  C. - Coulomb Potentials by Path Integration. - Fortschr.Phys., 1992, v.40,
	 p.695.
\item  Grosche C. - An Introduction into the Feynman Path Integral. -
	Preprint NTZ 29/1992; arXiv:hep-th/9302097.
\item Dulweber A., Hilf E.R., Mendel E. -
	Simple Quantum Mechanical Phenomena and the Feynman Real Time Path Integral
	Report-no: UO-PHYS-THEO 28 Nov. 1995 (Universit\"at Oldenburg); arXiv:quant-ph/9511042.
\item Mendel E., Nest M. - Time evolution for quantum systems at finite    temperature. -
	Nucl.Phys. B, 1999, v.562, p.567; 
	arXiv:hep-th/9807030.
\item  Mak C.H.,  Egger R. - 
	A Multilevel Blocking Approach to the Sign Problem in Real-Time Quantum
	Monte Carlo Simulations. - J. Chem. Phys., 1999, v.110, p.12;
	arXiv:physics/9810058.
\item Filinov V.S., Bonitz M., Ebeling W. , Fortov V.E. - Thermodynamics
	of hot dense H-plasmas: Path integral Monte Carlo simulations and analytical
	approximations. - Plasma Physics and Controlled Fusion, 2001, v.43, p.743.
\item Rejcek J.M, Datta S., Fazleev N.G., Fry J.L. - Application of the Feynman-Kac
	path integral method in finding excited states of quantum systems. -
	Computer Physics Communications, 2002, v.146, p.154. 
\item Bond S.D.,  Laird B.B., Leimkuhler B.J. -
	On the approximation of Feynman-Kac path integrals. -
	Journal of Computational Physics, 2003, v.185, p.472. 
\item Myrheim, J. - Numerical path integration with Coulomb potential. -
	arXiv:physics/0306168, 2003.
\end{enumerate}

}

\end{document}